\documentclass[epj]{webofc}
\usepackage[varg]{txfonts}   
\wocname{EPJ Web of Conferences}
\woctitle{CONF12}
\newcommand{\al}{\alpha}
\newcommand{\be}{\beta}
\newcommand{\ga}{\gamma}
\newcommand{\de}{\delta}
\newcommand{\la}{\lambda}

\newcommand{\NN}{{\cal N}}
\newcommand{\dott}[1]{\partial_{\,\ln\mu}{#1}}
\newcommand{\Gt}{\W_{\tau}}
\newcommand{\GaWZ}{S_{\rm WZ}}
\newcommand{\UV}{\rm UV}
\newcommand{\IR}{\rm IR}

\newcommand{\LUV}{\Lambda_{\UV}}
\newcommand{\Qmu}[1]{[{\mathcal{D}} #1]}
\newcommand{\ZZ}{\hat{ Z}}
\newcommand{\tr}{{\rm tr}}

\newcommand{\Dt}[1]{ D^{(#1)} }
\newcommand{\Wal}{\al}
\newcommand{\gQCD}{g }
\newcommand{\gm}{\mathrm{g}}

\newcommand{\SW}{S_{\rm W}}
\newcommand{\W}{W}
\newcommand{\ff}{s}
\newcommand{\ba}{a}
\newcommand{\bb}{b}
\newcommand{\bc}{c}

\newcommand{\Rtr}[2]{\Theta}

\newcommand{\bcp}{c'}
\newcommand{\baS}{a^{\rm free}_{(0)}}

\newcommand{\vev}[1]{\langle #1 \rangle} 

\newcommand{\til}[1]{\tilde{#1}}

\begin{document}
\selectlanguage{english}
\title{${\cal N}$ = $1$ Euler Anomaly  from RG-dependent metric-Background}
%
%

\author{Vladimir Prochazka\inst{1} \and
        Roman Zwicky\inst{1} \fnsep\thanks{\email{roman.zwicky@ed.ac.uk} -- Preprint number: 
        CP3-Origins-2016-053 DNRF90}
  }

\institute{Higgs Centre for Theoretical Physics, School of Physics and Astronomy,\\
University of Edinburgh, Edinburgh EH9 3JZ, Scotland, UK }

\abstract{%
 We consider ${\cal N}=1$ supersymmetric gauge theories in the conformal window. 
 By applying a suitable matter superfield rescaling and a
Weyl-transformation the renormalisation group running (matter and gauge field $Z$-factors) are absorbed into the metric. The latter becomes a function of the $Z$-factors. 
The  Euler flow 
$\Delta a \equiv a_{\rm UV} - a_{\rm IR} |_{{\cal N}=1}$
is then obtained by  free field theory computation with the non-trivial dynamics coming 
from expanding the Euler invariant in the flow dependent metric. The result is 
therefore directly obtained in terms of the infrared anomalous dimension
confirming an earlier result using the matching of  conserved currents.}
\maketitle
\section{Introduction}
\label{intro}

For a theory on a curved space in 4D,  with no explicit scale symmetry breaking, the 
trace of the energy momentum tensor TEMT is parametrised by \cite{CD78,20years}
\begin{equation} 
\label{eq:TA}
  \vev{ \Rtr{T}{\rho} } =  \ba \, E_4   + \bb\, W^2 + \bc \, H^2 + \bcp \Box H \;,    \quad \quad H \equiv \frac{R}{d-1} \;,
\end{equation}
where $E_4$, $W^2$ and $R$ are the topological Euler term, the Weyl tensor squared 
and the Ricci scalar respectively. The Euler anomaly has a long history in relation 
to the a-theorem, e.g. \cite{C88,JO90,KS11} and \cite{S16} for a review. 
The theorem states that  
the difference of the ultraviolet (UV) and infrared (IR)  Euler anomaly
$\Delta a \equiv a_{\UV} - a_{\IR} $ is strictly positive and intuitively 
measures the loss of degrees of freedom during the flow.
 For  $\NN=1$ supersymmetric gauge theories an exact expression was derived, valid in the conformal window, \cite{AFGJ97} almost twenty years ago.
Ever since it has served as  a fruitful laboratory for testing  different techniques by rederiving 
 the result. Examples include  four-loop perturbation theory  \cite{JP14},   the local renormalisation group (RG)   \cite{BKZR14} and employing  
 superspace techniques  
 assuming  a gradient flow equation \cite{AKZ15}
 conforming an older conjecture  \cite{FO98}.  
 
Here we derive $\Delta  \ba |_{\NN=1}$ by absorbing the matter and gauge 
$Z$-factors into the background metric. 
This renders the theory, in the vacuum sector, equivalent to 
a free field theory in a curved background carrying the information of the dynamics 
$ \til{\gm}_{\rho \la}  = f( Z,\ga_*) \de_{\rho \la}$. 
The difference of the Euler anomaly is computed 
using the conformal anomaly matching and dilaton effective action 
techniques used by Komargodski and Schwimmer  (KS) \cite{KS11,K11} to prove 
the a-theorem but at the same time differs substantially from it.

\section{Framework}
\label{sec-1}
In this section we derive the formula for computing $\Delta \ba$.
Let us consider 
a massless theory with a generic fields  $\phi$ and  
 coupling $g$. The IR effective $W$ is given by  
\begin{equation}
\label{eq:partition}
 e^{ \W(g(\mu),\mu) }   =   \int \Qmu{\phi}_\mu \ e^{- \SW( g(\mu),\mu,\phi)}  \;,
\end{equation}
where the subscript $W$ on the action stands for Wilsonian. Hence $S_W$ is on 
a renormalisation trajectory from the UV to the IR fixed point. A generic Weyl rescaling 
of the metric $\gm_{\mu \nu} \to e^{-2 \Wal(x)} \gm_{\mu \nu}$ results in 
$q^2 \to q^2 e^{2 \Wal(x)}$ and since the correlation functions depend on ratios of 
$q^2 /\mu^2$ the operation is equivalent to $\mu \to e^{ - \Wal(x)} \mu$. 
The key idea in \cite{KS11} is to compensate for this change by coupling  
a dilaton to the RG scale $\mu e^{\tau(x)}$ which transforms as $\tau(x) \to \tau(x) +  
\Wal(x)$.  The action in \eqref{eq:partition} then assumes the form 
$ \SW(g(\mu), \mu, \phi) \to    \SW^{(\tau)}(g(\mu e^{ \tau}), \mu e^{ \tau}, \phi )$ 
and is formally dilation invariant.
The partition function  assumes the form 
\begin{equation} 
\label{eq:taupath}
 e^{   \W_\tau( g(\mu),\mu,\tau(x)) } =   \int \Qmu{\phi}_\mu
 e^{- \SW^{(\tau)}(g(\mu e^{ \tau}), \mu e^{ \tau}, \phi ) }  \;,
\end{equation}
and becomes a function of the \emph{external} dilaton field. The dilaton is the spurion of the RG transformation and its response (effective action) must be sensitive to the difference in UV and IR 
conformal anomalies. 
It was shown in \cite{KS11} that the relevant term is
\begin{equation}
\label{eq:2}
\Gt( g^*_{\UV}) - \Gt(g^*_{\IR})  =  \int_{g^*_{\IR} }^{g^*_{\UV} } d g  \, \partial_g {\W}_\tau
 = -   \int_{-\infty}^{\infty} d \ln \mu \, \dott{\W}_\tau 
 = - \Delta \ba \ \GaWZ + \dots \;,
\end{equation}
where the Wess-Zumino action is given by 
\begin{equation}
\label{eq:gaWZ}
\GaWZ =   \int d^4 x \, 2 \left(  2\Box \tau (\partial \tau)^2- (\partial \tau)^4 \right) + {\cal O}(R)  \;.
\end{equation}
and since $\dott{\W}_\tau  =   \int d^4 x  \sqrt{ \tilde \gm} \ \vev{ \Rtr{{T}}{\rho}}_{\tau} $ (e.g. appendix \cite{Prochazka:2015edz}) 
one arrives at the key formula 
\begin{equation}
\label{eq:4}
\Delta \ba =   \int_{-\infty}^{\infty} d\ln \mu \ 
\int d^4 x \sqrt{ \tilde \gm} \vev{ \Rtr{{T}}{\rho}}_{\tau}|_{\GaWZ} \,,
\end{equation}
where $|_{\GaWZ}$ denotes  projection and  $\vev{ \Rtr{{T}}{\rho}  }_\tau $ is 
the TEMT in the dilaton background. Assuming a conformally flat metric 
$\til{\gm}_{\rho \la} =  e^{-2\ff(\tau)} \de_{\rho \la} $  the latter 
reads 
\begin{equation} 
\label{eq:TA-ac}
  \vev{ \Rtr{{T}}{\rho}  }_\tau  = \til{\ba} \til{E}_{4} + \til{\bc} \til{H}^2  + 
  \til{c}' \Box H  = 
  \vev{ \Rtr{{T}}{\rho}}_{\tau}|_{\GaWZ}  \, \GaWZ  + ... \;,
  \end{equation}
  where it was used that the Weyl tensor vanishes in those spaces since 
  it is sensitive to the spin 2 part only and we discard the $\Box \til{H}$-term since 
  being a total derivative it does not contribute $\GaWZ$. 
Hence determining $\Delta \ba$ reduces to finding $\til{\ba}$ and $\til{\bc}$ which of course 
depend on the content of the theory.  
Before turning to the $\NN=1$-computation it is instructive and useful to first 
discuss the case of a single scalar field.

\subsection{The scalar field with dynamics absorbed in the metric as an example}
\label{sec:toy}

Consider a scalar field theory in  flat space with Wilsonian action 
\begin{equation}
\label{eq:S0}
\SW(\mu) =  \int d^4 x Z(\mu)  \de^{\rho \la}  \partial_{\rho} \phi  
\partial_{\la} \phi  + \dots\;,
\end{equation}
where it is going to be sufficient, for later purposes,  to focus on the kinetic term only.
The dilaton is implemented by  $Z(\mu) \to Z(\mu e^{\tau(x)})$ 
and it is the key idea of this paper that the entire RG running 
$Z(\mu e^{\tau(x)})$ can be absorbed into metric by the Weyl rescaling 
with $\al(x) = s(x)$ given by 
\begin{equation}
\label{eq:g-t}
  \ff(\mu e^\tau(x)) = -\frac{1}{2} \ln Z(\mu e^{\tau(x)}) \quad  
\Rightarrow  \quad
   \til{\gm}_{\rho \la}  =  e^{-2 s(x)}  \de_{\rho \la} =   Z  \de_{\rho \la} \;.
\end{equation}
The theory then  becomes a free field theory $S^{(\tau)}_W(\mu) =  \int d^4 x \sqrt{\til{\gm}}  \til{\gm}^{\rho \la }  
\Dt{\ff}_{\rho} \phi \Dt{\ff}_{\la} \phi \;$
on  a conformally flat space  
with metric $\til{\gm}_{\mu \nu}$ \eqref{eq:g-t}. In the previous formula  
$\Dt{\ff}_\rho = \partial_\rho -(\partial_\rho \ff)$ are covariant derivatives ensuring 
local Weyl invariance, a technical detail for which we refer the reader 
to \cite{Prochazka:2015edz}. 

To get the Euler anomaly we then need 
to determine \eqref{eq:TA-ac} in a free field theory and expand the corresponding 
geometric term which will involve the  dynamics in terms of the anomalous dimension 
(appearing through derivatives of the absorbed $Z$-factor). The TEMT in a free theory  
 (cf.  \cite{CD78,birrell1982quantum} or the explicit computation in appendix  
 \cite{Prochazka:2015edz})  is given by
 \begin{equation}
\label{eq:Afree0}
\vev{ \Rtr{{T}}{\rho}  }_\tau =
\baS  ( \til{E}_4 -2 \til{\Box} \til{R})
  \;, \qquad \baS =
  \frac{1}{ 360} \frac{1}{16 \pi^2} = \frac{1}{5760 \pi^2} \;,
\end{equation}
where the absence of the $R^2$-term is a consequence of the free field theory 
being a conformal field theory. As previously explained 
the $\Box R$-term  does not contribute to $\GaWZ$ and can therefore be discarded. 
Evaluating  the latter in the background  \eqref{eq:g-t} leads to
\begin{eqnarray}
\label{eq:E4}
\!\!\!\!\! \!\!\!\!\!   \sqrt{\til{\gm}} \til{E}_4 &=&
- 8 ( \frac{1}{2}\Box (\partial \ff)^2 - \partial \cdot(\partial  
\ff(\Box \ff - (\partial \ff)^2)))   \\[0.1cm]
&=& -[  \ga^2  \Box (\partial \tau)^2 +(2\ga  
\dot{\ga} - 2\ga^2)  \partial^{\la}(\partial_{\la} \tau \Box  
\tau) - \ga^3  \partial^{\la}(\partial_{\la} \tau (\partial  
\tau)^2)  - 
 6\ga \dot{\ga} (\partial \tau)^2 \Box \tau
-3 \ga^2 \dot{\ga} (\partial \tau)^4 \big ] \;,\nonumber
\end{eqnarray}
 where  the following abbreviation  $\dot{\ga} \equiv  
\frac{d}{d \log \mu}{\ga}$ was used as well as
\begin{equation} \label{TildeDerivative}
\partial_\rho \ga =  \dot{\ga} \, \partial_{\rho}\tau   \;,  \quad
  \partial_{\rho} \ff =  - \frac{1}{2} \frac{\partial \ln Z(\mu  
e^{\tau}) }{\partial (\mu e^{\tau})}
  \partial_\rho (\mu e^{\tau} )  =  - \frac{1}{2} \ga \,  
\partial_{\rho}\tau \;, \quad
\ga   =  \frac{\partial \ln Z(\mu) }{\partial \ln \mu}  \;.
\end{equation}
 Note that 
  $\ga$ and $\dot{\ga}$ can be treated as being space-independent,
since expanding  $\gamma(\mu e^{\tau})= \gamma(\mu)+ O(\tau(x))$ leads  
to terms which are not in  $\GaWZ$ \eqref{eq:gaWZ}.
Furthermore  it is then clear that the total derivative terms in  \eqref{eq:E4}  
can be discarded since they 
do not contribute to the bulk-term  $\GaWZ$ \eqref{eq:gaWZ}.  
 In order to project on $\GaWZ$
it is convenient (following \cite{KS11},\cite{K11}) to set $\Box \tau = (\partial \tau)^2$
 under which  
 $\GaWZ \to  \int d^4 x  \, 2 (\Box \tau)^2 $. Using \eqref{eq:4} one gets the simple result
 \begin{equation}
\label{eq:Da-free}
\Delta \ba    
=   \frac{1}{2}\left(
(\ga_{\UV}^3 - \ga_{\IR}^3)  + 3 (\ga_{\UV}^2 - \ga_{\IR}^2) \right) \baS \;,
\end{equation}
from the last two terms in \eqref{eq:E4}. Above   
 $\ga_{\IR,\UV} \equiv \ga( g^*_{\IR,\UV}) $ are the values of the anomalous dimensions at the respective fixed points.
 Eq.~\eqref{eq:Da-free} constitutes an important intermediate result for the derivation of $\Delta a|_{ \NN=1}$.

\section{$\NN = 1$ supersymmetric gauge theory}
\label{sec:N=1}

We aim to compute $\Delta \ba$ for an $\NN=1$ supersymmetric gauge theory with  $SU(N_f) \times SU(N_f)$ flavour symmetry and gauge group $SU(N_c)$. 
In the superfield formalism the  action  takes the form, e.g. \cite{S-book},  
\begin{equation}
\label{eq:S,N=1}
 \SW(\mu) = \int d^6 z \frac{1}{\gQCD^2(\mu)} \tr W^2 + {\rm h.c.} + 
 \frac{1}{8} Z(\mu) \sum_{f} \big [\int d^8 z    \Phi_{f}^\dagger e^{-2 V} \Phi_{f} + \{ \tilde{\Phi}_{f}  
 \leftrightarrow \Phi_f  \}   \big ]\;,
\end{equation}
with vector  $V$ and matter $(\Phi_{f},\tilde{\Phi}_f)$ superfields. Above  
$W^2$ is the supersymmetric gauge field kinetic term (function of $V$),  $\gQCD$ is  the holomorphic  coupling constant  and 
$d^6 z $ and $d^8 z$ include integration over the fermionic superspace variables. 
The main tool in deriving $\Delta  \ba |_{\NN=1}$ is the use of the Konishi anomaly \cite{K83,KS85,ShV86} which is also of use in deriving the NSVZ $\be$-function e.g. 
\cite{S-book} or the appendix of \cite{Prochazka:2015edz}. Using the Konishi anomaly, 
and a suitable Weyl rescaling  
the action \eqref{eq:S,N=1} can be cast into a form that makes it possible to use 
the  previously studied free field theory  example.

\subsection{$\Delta  \ba |_{\NN=1}$ from Dilaton effective Action and Konishi anomaly }
\label{sec:N=1,comp}

Using arguments of holomorphicity it can be argued that the running of the coupling $\gQCD$, 
of the Wilsonian effective action of the supersymmetric gauge theory  \eqref{eq:S,N=1}, 
is one-loop exact  \cite{AM97,S-book} and reads
\begin{eqnarray}
\label{eq:Seff,N=1}
 \SW(\mu) = \frac{Z_{1/g^2}(\mu,\mu')}{\gQCD^2(\mu')}   \int d^6 z  \tr W^2 + {\rm h.c.} +  \frac{1}{8} Z(\mu,\mu') \sum_{f} \big [\int d^8 z    \Phi_{f}^\dagger e^{-2 V} \Phi_{f} + \{ \tilde{\Phi}_{f}  
 \leftrightarrow \Phi_f  \}  ]\;,
\end{eqnarray}
where 
\begin{equation}
Z_{1/g^2} = 1+ \ga_* N_f \frac{\gQCD(\mu')^2}{8 \pi^2} \ln \frac{\mu'}{\mu}  \;,  
\quad \ga_* N_f = -b_0 = -(3N_c- N_f) \;,
\end{equation}
is the $Z$-factor of the $1/\gQCD^2$-coupling, $\ga_*$ the IR squark anomalous dimension 
 and the arbitrary scale $\mu' > \mu$  can be identified 
with the UV cut-off $\LUV$.
We emphasise that $ \SW(\mu) $ is $\mu'$-independent.
The Konishi anomaly\footnote{The Konishi anomaly accounts 
for the rescaling of individual field, in a regularisation with a cut-off, and therefore 
contributes to the analogue of the field strength term (i.e. $W^2$) as dictated by the trace anomaly. 
It is not peculiar to supersymmetry but the power within supersymmetry is that it is one-loop
exact since it is in the same multiplet as the chiral anomaly and therefore inherits its topological protection.}
 allows, by rescaling the matter fields    
\begin{equation}
\label{eq:rescaling2}
(\Phi_{f},\tilde{\Phi}_f) \to  \left( \frac{\mu'}{\mu}\right) ^{\ga_*/2} (\Phi_{f},\tilde{\Phi}_f) \;,
\end{equation}
to shift the  $Z_{1/g^2}$-factor in front of the matter term
\begin{eqnarray}
\label{eq:Sbare}
 \SW(\mu) &\;=\;& \int d^6 z \frac{1}{\gQCD(\mu')^2}  \tr W^2 + {\rm h.c.} +  \frac{1}{8} \sum_{f} \big [\int d^8 z \, \hat{Z}(\mu,\mu')   \Phi_{f}^\dagger e^{-2 V} \Phi_{f} +   \{ \tilde{\Phi}_{f}  
 \leftrightarrow \Phi_f  \}  \big ]\;,
\end{eqnarray}
with
\begin{equation}
\ZZ (\mu ,\mu') \equiv Z(\mu,\mu') \left( \frac{\mu'}{\mu} \right)^{\ga_*} \;,
\end{equation}
where the arbitrary scale  $\mu' > \mu$  can be interpreted as  a UV cut-off $\LUV$. 
Crucially, the entire RG flow is absorbed into the precoefficient  $\ZZ (\mu ,\mu')  $ in front of the matter term. 
Eq.~\eqref{eq:Sbare} is the analogue of the action \eqref{eq:S0} for the scalar field to the degree
that the running of the theory is parametrised by a coefficient in front of the matter kinetic term.
The external dilaton field   is introduced as before 
\begin{equation}
\ZZ(\mu,\mu')  \to  \ZZ(\mu e^{\tau(x)},\mu') =
Z(\mu e^{\tau(x)},\mu') \left( \frac{\mu'}{\mu e^{\tau(x)}} \right)^{\ga_*}  \;.
\end{equation}

The crucial point is that the $W^2$-term is Weyl invariant where the matter term 
is Weyl variant which allows to absorb the dynamics into the metric, i.e. $\ZZ(\mu,\mu')$. 
This is achieved by 
the following Weyl rescaling $\alpha = s$\footnote{The  procedure can be kept 
manifestly supersymmetric, following \cite{ST10},  by promoting the dilaton to a (chiral) superfield $T$ 
and a superfield $A$ playing the role of the supersymmetric Weyl parameter. 
We refer to  \cite{Prochazka:2015edz} for details.}
\begin{equation}
\label{eq:Ztil}
\til{\gm}_{\rho \la} = \til{e}^a_\rho  \til{e}^a_\la = e^{-2\ff(\mu e^\tau(x))}  \de_{\rho \la} \;, 
\quad \ff(\mu e^\tau(x)) = -\frac{1}{2} \ln \ZZ(\mu e^{\tau(x)},\mu') \;,
\end{equation}
where $\til{e}^a_\rho$ is the vielbein and  $\til{\gm}_{\rho \la}$ is the  background carrying the dynamics 
and the remaining part is a free field theory.\footnote{Note this is only true in the vacuum sector as the introduction of correlation function in terms of sources would involve further 
changes under the Weyl-transformation.}

Since the UV scale $\mu'$ is arbitrary the  physical quantity   $ \vev{ \Rtr{{T}}{\rho}  }_\tau$  is independent of it and we may therefore send it to infinity. 
Since the geometric terms   $\til{E}_{4}$ and 
$\til{H}^2$ are independent of $\mu'$,\footnote{To see this  note that these terms  depend on derivatives of $s$ only (c.f \eqref{eq:E4}). The latter are related to the anomalous dimension $\gamma(\mu e^{\tau})$ through the relation \eqref{TildeDerivative} 
which is independent of $\mu'$.}  the form of \eqref{eq:TA-ac} 
implies that $\til{a}$ and $\til{c}$  are $\mu'$-independent and  therefore constants.  
This means that $\til{a}$ and $\til{c}$ take on the values at the (free)  UV fixed point 
 and the geometric quantities are to be evaluated using the background metric $\til{g}_{\rho \la}$. This allows us to largely reuse the computation in   section  \ref{sec:toy} as outlined below.

 The steps in completing 
 the computation are as follows.
Equivalence to the  example in the previous section is attained by  
replacing  $Z \to \hat{Z}$ in Eqs. \eqref{eq:g-t} to \eqref{eq:Da-free}, (following from \eqref{eq:Ztil}) implying  
$\ga \to \de \ga \equiv \ga - \ga_*\,$ \emph{and} 
accounting for the correct number of degrees of freedom $\nu$.  
Note that only matter fields contribute to the latter since $\hat{Z}$ only stands in front 
of the matter term.  The matter superfield consists of a complex scalar and a Weyl fermion which contribute 
 \cite{CD78}
\begin{equation}
\nu \equiv 2\Big|_{\mathbb{C}\text{-scalar}} + \frac{11}{2}\Big|_{\text{Weyl-fermion}}   = \frac{15}{2}
\end{equation}
 in units of a real scalar field. This number has to be   multiplied by 
by flavour and colour numbers:  $2 N_f$ (two matter-field per flavour) and $N_c $ (the $SU(N_c)$ Casimir of 
the adjoint representation).  
Finally  $\Delta a$ is given 
by $ 2 N_f N_c\nu \Delta a|^{\eqref{eq:Da-free}}_{\ga_{\UV,\IR}  \to  \de \ga_{\UV,\IR}}$. 
With  $(\ga_{\UV},\ga_{\IR}) =  (0,\ga_*)$ leads to $(\de \ga_{\UV},\de \ga_{\IR})  = (- \ga_*,0)$   and finally  
\begin{equation}
\label{eq:N=1}
\Delta  \ba |_{\NN=1} =  \frac{15}{2}  N_c N_f (- \ga_*^3  + 3 \ga_*^2) \baS \;.
\end{equation}
The result \eqref{eq:N=1}  is indeed the same as the non-perturbative result 
quoted in (Eq.4.18) in \cite{AFGJ97} when taking into account the explicit form 
of  $\ga_*$.  
The formula \eqref{eq:N=1} is valid  in the conformal window 
$3/2 N_c < N_f < 3 N_c$  where the  theory is asymptotically free in the UV 
and the  acquires a non-trivial fixed point in the IR.  
Within these boundaries the anomalous dimension 
$\ga_*$ takes on the values $-1$ to $0$ and  $\Delta a $ is manifestly positive 
in accordance with the a-theorem.  
  Within $\NN=1$ supersymmetric theories the a-theorem has many applications 
  including the so-called $a$-maximization \cite{IW03}.

\section{Discussion}

The main result of this work is the computation of $\Delta  \ba |_{\NN=1}$ 
\eqref{eq:N=1} in the new framework where the dynamics, i.e. 
the $Z$-factor and  $\ga_*$ the anomalous dimension at the IR fixed point,  
is absorbed into the metric \eqref{eq:Ztil}. 
This  renders the computation 
equivalent to the Weyl anomaly of a free field theory with non-trivial dynamics 
 originating from expanding the curvature term $E_4$ in terms of the metric \eqref{eq:Ztil}.
A possible advantage of our approach over the original derivation \cite{AFGJ97} is that 
the result is directly obtained in terms of the anomalous dimension of the squarks 
at the IR fixed point rather than identifying the latter indirectly from a result in terms 
of $N_C$ and $N_f$.
Let us conclude by saying that the procedure is reminiscent of the AdS/CFT set-up 
in that dynamics is encoded in the metric. \\

RZ would like to thank the  Confinement'16 organisers  and particpants   for a memorable event in Thessaloniki.



%
%
%

\bibliography{input}
\bibliographystyle{utphys}


\end{document}